\documentclass[preprints,review,accept,moreauthors,pdftex]{Definitions/mdpi} 
\firstpage{1} 
\pubvolume{7}
\issuenum{11}
\articlenumber{397}
\pubyear{2021}
\copyrightyear{2021}
\externaleditor{Academic Editor: {Maria Vasileiou}} % MDPI:please confirm if this should be added. For journal Automation, please change Academic Editor to "Communicated by"
\datereceived{30 September 2021} 
\dateaccepted{17 October 2021} 
\datepublished{21 October 2021} 
\hreflink{https://doi.org/\linebreak 10.3390/universe7110397} % If needed use \linebreak
\Title{Multimessenger Astronomy with Neutrinos}

\TitleCitation{Multimessenger Astronomy with Neutrinos}

\Author{Francisco %	MDPI:Please carefully check the accuracy of names and affiliations. Changes will not be possible after proofreading.
 Salesa Greus $^{1,2,}$*\orcidA{} and Agust\'{i}n S\'{a}nchez Losa $^{1}$\orcidB{}}

\AuthorNames{Francisco Salesa Greus and Agust\'{i}n S\'{a}nchez Losa}

\AuthorCitation{{Salesa Greus, F.; \linebreak S\'{a}nchez Losa, A.}} 
\address{%
$^{1}$ \quad IFIC-Instituto de Física Corpuscular, (CSIC---Universitat de València), c/ Catedrático José Beltrán, 2 E-46980 Paterna, Spain; Agustin.Sanchez@ific.uv.es\\
$^{2}$ \quad Institute of Nuclear Physics, Polish Academy of Sciences, PL-31342 Krak\'{o}w, Poland}

\corres{Correspondence: sagreus@ific.uv.es; Tel.: +34-963543538}

\abstract{Multimessenger astronomy is arguably the branch of the astroparticle physics field that has seen the most significant developments in recent years. 
In this manuscript, we will review the state-of-the-art, the recent observations, and the prospects and challenges for the near future. We will give special emphasis to the observation carried out with neutrino telescopes.}


\begin{document}
%%%%%%%%%%%%%%%%%%%%%%%%%%%%%%%%%%%%%%%%%%
\section{Introduction}
Astronomical observations have been traditionally done with visible light. From~the times of Galileo times until now we have been able to expand the observation range to all the electromagnetic spectrum, from~radio to gamma rays.
Apart from photons, the~discovery of other particles has opened the possibility of using them as cosmic messengers to explore the Universe. For~instance, cosmic rays (CRs) were first discovered in the early 1910s. We know that CRs are ionized nuclei of extraterrestrial origin, and~that they are produced and accelerated in a broad energy range, reaching energies above 10$^{20}$ eV. The~spectrum of the CRs detected at Earth decreases with the energy, following approximately a power law E$^{-n}$ with n$\sim$3. The~Sun is the main source of CRs below GeV energies. Up~to PeV energies, CR are believed to be dominated by Galactic sources, and~then, at~the highest energies, CRs are likely of extragalactic origin~\cite{ParticleDataGroup:2020ssz}. As CRs are charged, their directionality is lost due to the Galactic magnetic field. This is probably one of the main reasons why the origin of the most energetic CRs is still unknown.
We have also detected neutrinos from extraterrestrial origin. First from the Sun~\cite{Davis:1968cp}, and~then from a nearby supernova explosion in 1987~\cite{Hirata:1988ad,IMB:1987klg,Baksan:1987} which can be considered as the birth of neutrino astronomy. By~that time, the first project to construct a neutrino telescope was already ongoing~\cite{DUMAND:1989dxw}. However, only after decades of research and development, neutrino astronomy had its turning point in 2014 with the discovery of a high-energy cosmic neutrino flux by the IceCube collaboration~\cite{IceCube:2014stg}.
The most recent cosmic messengers discovered are gravitational waves (GWs). The~path to the first GW detection was also not easy, and~it took a century from their theoretical prediction to the first confirmation in 2015 by the LIGO and VIRGO collaborations~\cite{LIGOScientific:2016aoc}.

%Both neutrinos and GWs are challenging to detect and are usually associated with very high energetic events.

Multimessenger astronomy originates as a consequence of astrophysical neutrino and GW detection techniques reaching maturity. Multimessenger astronomy is based on the observation of four cosmic messengers, namely, photons, CRs, neutrinos, and~GWs. The~detection in coincidence of all, or~some, of~these messengers allows the study of a source in a similar fashion as it has been done with multiwavelength electromagnetic observations. Moreover, the~fact of involving different types of particles adds extra information coming from interactions involving all fundamental forces of~nature.

We have divided this review into three sections. In~Section~\ref{Milestones}, we will discuss the two events that marked the start of the multimessenger era, four years ago. In~Section~\ref{Results}, we will review the most recent results and what we have learned from them. To~conclude, in~Section~\ref{Future}, we will discuss the future prospects and challenges in the~field.
 
%%%%%%%%%%%%%%%%%%%%%%%%%%%%%%%%%%%%%%%%%%
\section{Multimessenger Astronomy~Milestones}
\label{Milestones}
Excluding the solar neutrino detection, the~observation by chance of neutrinos coming from the supernova explosion SN1987A is the first astronomical event producing a multimessenger (photon--neutrino) coincidence. However, there are two main events, both happening in 2017, that really marked the birth of a new field in astrophysics, multimessenger~astronomy.

The first event was the result of two neutron stars merging into a black hole. This produced a GW (GW170817) that was detected by the LIGO and Virgo~\cite{LIGOScientific:2017vwq} Scientific Collaborations. Less than 2 seconds after the event, a~short gamma-ray burst (GRB) (GRB 170817A) was detected by the Fermi and INTEGRAL satellites. This coincidence triggered a campaign where several observatories followed-up the event in an unprecedented way~\cite{LIGOScientific:2017ync}.
Thanks to this coincident detection it was possible to determine the location and the type of sources involved, bringing the first experimental evidence of a kilonova~\cite{Metzger:2017}, a~type of transient event, theoretically predicted more than two decades ago~\cite{Lixin:1998}, where nucleosynthesis of the heavy elements is produced.
%Nucleosynthesis of the heavy elements (r-process) 

The second event (IC-170922A) was triggered by a high-energy neutrino, of~about 300 TeV, detected by the IceCube observatory on 22 September 2017. Again, this event was extensively followed up by other observatories. 
In this case, observations from the Fermi-LAT satellite were able to point out a blazar (TXS0506+056) in active state which was in spatial and temporal coincidence with the neutrino event. The~event was rejected to be produced by background fluctuations at 3$\sigma$ level~\cite{IceCube:2018dnn}.
After this detection, archival analysis of IceCube data prior to the IceCube-170922A event unveiled a potential flare in neutrinos~\cite{IceCube:2018cha}, between~September 2014 and March 2015, with~3.5$\sigma$ statistical significance and independent of the 2017 neutrino alert. In~this case, no gamma-ray counterpart was observed.
An extensive multiwavelength monitoring of TXS0506+056 started after the coincident event in September 2017 showed a low state emission except for December 1st and 3rd, 2018 with a flare comparable to the one in 2017~\cite{Satalecka:2021}. However, no neutrino excess was observed.
It is also important to mention that TXS0506+056 came out as the second most significant source (2.8$\sigma$ pre-trial) in the point source search analysis done with the ANTARES neutrino telescope using a pre-selected list of sources~\cite{Illuminati:2021b}, which makes the case of TXS0506+056~stronger.

%%%%%%%%%%%%%%%%%%%%%%%%%%%%%%%%%%%%%%%%%%
\section{Recent~Results}
\label{Results}
Once it has been well established that a flux of high-energy neutrinos of cosmic origin exist~\cite{IceCube:2014stg}, the~next step is to disentangle it and identify which are the sources. The~most recent all-sky searches performed by ANTARES~\cite{Illuminati:2021b} and IceCube~\cite{IceCube:2019cia} did not reveal any significant detection above the discovery threshold of 5$\sigma$, with the excess near the galaxy NGC 1068 observed by IceCube being the most interesting spot with a post-trial significance of 2.9$\sigma$.
This type of high-energy searches benefit from multimessenger astronomy thanks to including the sky coordinates and timing information from potential cosmic messenger counterparts. That was how the first evidence of a cosmic neutrino source, TXS0506+056, was~found. 

Understanding the multimessenger emission from TXS0506+056 has been challenging from the theoretical point of view, as it is difficult to get a good agreement between the observed neutrino signal and other wavelength observations. 
If one tries to explain it with a single-zone model, i.e.,~both gammas and neutrinos coming from the same region, one finds out that a leptonic scenario with a radiatively subdominant hadronic component provides the only physically consistent single-zone picture~\cite{Keivani:2018rnh}.
%Keivani, 2018 ApJ 864 84
%We find that a leptonic scenario with a radiatively
%subdominant hadronic component provides the only physically
%consistent single-zone picture for this source MM emission
%- Challenging source from the theoretical point of view. It is difficult to match the neutrino signal with other wavelengths observations (atypical photon luminosity required)
A higher neutrino flux would be expected if the source hosts two physically distinct emitting regions (see for instance~\cite{Xue:2019txw}). However, current observations cannot discriminate between single- or multi-zone emission models.
Related to this, a~compelling neutrino--radio correlation~\cite{Plavin:2020mkf} has been recently discovered. The~authors proposed that neutrinos and gamma rays may be indeed produced in different regions~\cite{Plavin:2020emb}. If~this is actually the case, X-ray and radio may be better wavelengths when looking for photon--neutrino correlations. 
The correlation with radio blazars is also supported by ANTARES observations~\cite{ANTARES:2020zng}.
In this regard, ANTARES has recently reported the results from an untriggered search from radio blazars with an interesting association coming from J0242+1101~\cite{Illuminati:2021}.

%%%%
Thanks to the IceCube alert system~\cite{IceCube:2016cqr}, which is presently providing on the order of 10 (20) gold (bronze) alerts per year\endnote{Gold (bronze) alerts are neutrino events with >50\% (>30\%) probability of being from astrophysical origin.}, more coincidences between neutrino and blazars have been found lately. For~instance, PKS 1502+106 blazar was coincident with a 300~TeV neutrino~\cite{Rodrigues:2020fbu}. In~this case, the blazar was in a quiescent state at the time of the neutrino alert. However, no more neutrinos were detected. 
Another example is 3HSP J095507.9, which was also coincident with a high-energy neutrino detected by IceCube~\cite{Paliya:2020mqm,Giommi:2020viy}. However, for~this event a lot of sources lay around the best position provided by IceCube, preventing the identification of a potential source candidate. This also underscores that sub-degree angular resolution, achievable by future neutrino observatories, will be key when looking for spatial coincidences.
From Fermi-LAT observations we know that blazars are the most abundant extragalactic gamma-ray sources, constituting roughly 80\% of the entire extragalactic source population~\cite{Fermi-LAT:2019pir}. However, current predictions based on stacking catalog searches performed with IceCube data estimate that neutrinos emitted by blazars, in~the range between around 10 TeV and 2 PeV, can only contribute up to 27\% to the total neutrino diffuse flux~\cite{IceCube:2016qvd}. 
More multimessenger observations with next generation experiments are required to test current theoretical models, and~therefore provide valuable information to understand the particle production and acceleration in~blazars.   

Apart from blazars other neutrino candidates have been already identified thanks to multimessenger observations. This is the case of Tidal Disruption Events (TDEs), which are the result of a star being ripped apart when passing next to a supermassive black hole. TDEs were already hypothesized as possible neutrino sources, e.g., in~\cite{Wang:2011}. However, it was not until 2019 when an IceCube neutrino, with~59\% probability of being of astrophysical origin (IC-191001A), triggered an alert that was followed-up by the Zwicky Transient Facility~\cite{Bellm:2019}. In~spatial coincidence with the neutrino alert a TDE (AT2019dsg) was observed. Given that TDEs are rare events, the~chance probability of finding this coincident event was estimated to be less than 0.5\%~\cite{Stein:2020xhk}.
%We can include here what are the suggested neutrino production zones + REFs
After the first neutrino-TDE coincidence, more recently another possible association has been detected (IC200530A with AT2019fdr) which has been considered by some scientists as evidence of an emerging trend. The~chance probability of finding this second event in coincidence was also small.
Both events have been followed up by ANTARES, however did not produce any significant neutrino {excess}~\cite{ANTARES:2021jmp}.~%F.S.G.:reference replaced
The~non detection does not contradict the observation by IceCube, as~the sensitivity of ANTARES was above the neutrino flux prediction.
On the other hand, there are preliminary indications of an excess in GVD-Baikal data~\cite{Allakhverdyan:2021}. %There is no corresponding bibitem for the citation 'Allakhverdyan:2021'
Current estimations predict that TDEs contribute at least 2\% but not more than 40\% of the total neutrino flux~\cite{Stein:2021}. Again, more observations are needed to confirm TDEs as sources of high-energy neutrinos and~determine their precise contribution to the diffuse neutrino~flux.

GRBs are another type of sources that have long been predicted as good candidates to emit high-energy neutrinos, see, for instance, in~\cite{Waxman:1997ti}. However, so far, all the searches have been unsuccessful~\cite{Albert:2016eyr,IceCube:2016ipa}.
Recently, some studies have tried to infer what would be the relative contribution of the different neutrino candidate sources, see, e.g., in~\cite{Bartos:2021tok}. However, it seems that there is no clear indication of a dominant type of source producing cosmic neutrinos. Interestingly enough, the~same study suggest that there is room for unknown~sources.

%%%%%%%%%%%%%%%%%%%%%%%%%%%%%%%%%%%%%%%%%%
\section{Future Prospects and~Challenges}
\label{Future}
\unskip
\subsection{Future Instruments and Instrument~Upgrades}

Regarding neutrino telescopes there are three major projects that will be operating in the near future.
KM3NeT~\cite{KM3Net:2016zxf} is a research infrastructure being built in the Mediterranean sea. KM3NeT is composed of two detectors, first ARCA (Astroparticle Research with Cosmics in the Abyss) which is designed to be sensitive to high-energy neutrinos in the TeV-PeV range, and~therefore with astrophysics studies as the main goal. The~second detector is called ORCA (Oscillation Research with Cosmics in the Abyss) and it is sensitive to GeV neutrinos. The~ORCA detector will primarily be used for the study of neutrino properties.
Both instruments will use the same technology and detection principle, i.e.,~array of photomultipliers tubes (PMTs) in sea water, being the main difference the volume covered, and~therefore the PMT density.
ARCA is expected to be fully operational in 2027 and ORCA in~2025.

Moreover, in water there is the GVD-Baikal project that is in construction in the Baikal Lake in Russia. GVD-Baikal is currently operational with 2304 optical modules arranged in eight~clusters of eight strings each. In~the present configuration, it has an effective volume of 0.4~km$^{3}$ for cascades with energy above 100 TeV~\cite{Baikal:2021}. Current plans are to deploy six~additional clusters for the period from 2022 to 2024 which should provide an additional 0.3~km$^{3}$ effective volume. %https://pos.sissa.it/395/002/pdf

The leading project in neutrino telescopes in the last decade has been IceCube~\cite{IceCube:2021} which is a neutrino telescope installed in the South Pole.
After 10 years of successful operation there are plans for two major upgrades. One, called IceCube-Gen2~\cite{IceCube-Gen2:2020qha}, expected to be completed by the early 2030s, will significantly increase the IceCube effective volume and energy range sensitivity. The~other, called IceCube-Upgrade~\cite{IceCube:2019xdf}, represents a fraction of a larger project called Precision IceCube Next Generation Upgrade (PINGU)~\cite{IceCube:2016xxt}, which aims to increase the sensitivity to lower energies even more than with IceCube-DeepCore, mainly thanks to a higher string density, and~that will mostly study neutrino~properties.

We can also add to the list of future intended neutrino telescopes the Pacific Ocean Neutrino Experiment (P-ONE)~\cite{PONE:2021}. The~P-ONE project is presently in research and development phase. The~goal of the collaboration is to install a multi-cubic-kilometer neutrino telescope in the Pacific Ocean, which is expected to be operational in the next~decade.

\textcolor{black}{Other neutrino experiments, foreseen to be operational by the end of this decade, like Hyper-Kamiokande~\cite{Hyper-Kamiokande:2018ofw} and DUNE~\cite{DUNE:2020lwj}, will be sensitive to lower neutrino energies (MeV-GeV). However, they can still contribute to get the whole picture of some astrophysical events. For~instance, the~explosion of a nearby supernova, where low-energy neutrinos are expected to be produced.}

\begin{figure}[H]
\includegraphics[width=0.95\linewidth]{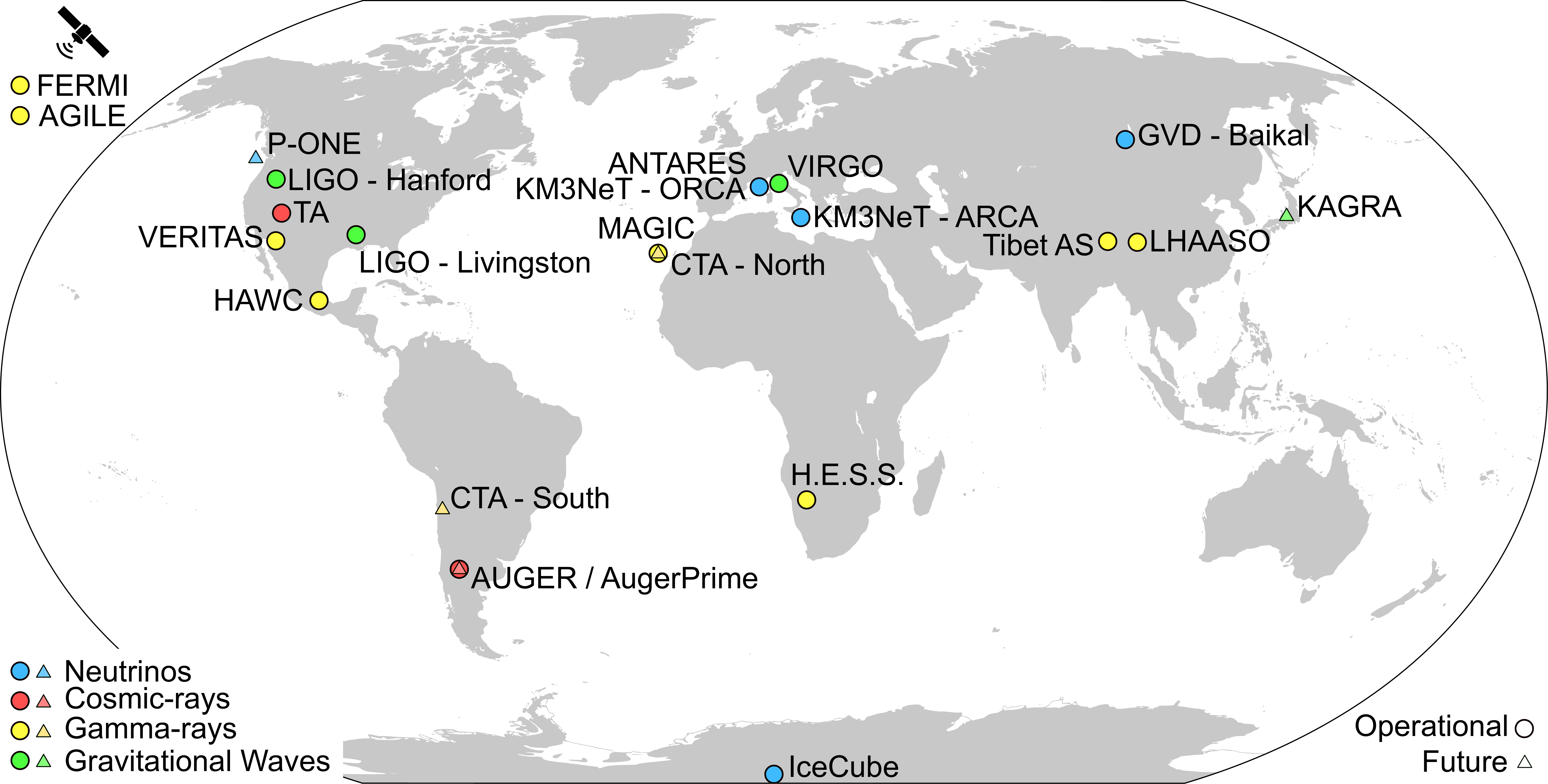}
\caption{Earth map indicating the location of a selection of multimessenger observatories that are either currently operating (circles) or planned (triangles). Satellites are depicted outside to the~map.}
\label{fig1}
\end{figure}

There are also very exciting plans for next generation experiments aiming to detect other cosmic messengers. Just to mention a few of them, we have the Cherenkov Telescope Array (CTA)~\cite{CTA:2021} with two planned sites (Northern and Southern Hemisphere), and~LHAASO~\cite{LHAASO:2021}, fully operational since July 2021, detecting gamma rays. KAGRA~\cite{KAGRA:2019}, detecting GWs, will join LIGO and Virgo for the next GW data taking run (O4), which is expected to start in late 2022. Finally, detecting ultra-high-energy CRs, there will be AugerPrime~\cite{PierreAuger:2016qzd}, the~upgrade of the Pierre Auger Observatory.
Such a network of observatories, distributed in different locations around the globe (see Figure~\ref{fig1}), will provide a full multimessenger coverage of the~sky.

\subsection{Alert Systems and~Strategies}

Considering the number of experiments currently being under construction and planned, it seems clear that an efficient communication between collaborations is crucial. Moreover, for~the case of pointing instruments, like CTA, this communication also needs to be fast. 
\textcolor{black}{Neutrinos are actually very good messengers to trigger alerts because they are able to easily escape from sources. These neutrino alerts can give an early warning to other observatories of an incoming event, allowing for a prompt follow up of transient~phenomena.}

There are already ways to announce, in~real-time, interesting events to the astrophysics community, e.g., the~GammaRay Coordinates Network (GCN)~\cite{GCN:2021}. In~addition to these announcements, there are also sites, like the Astronomers Telegram (ATEL)~\cite{ATEL:2021}, where a brief report about recent observations made by the experiments is posted online.
Still in beta testing phase there is a project called {Astro-COLIBRI}~\cite{Reichherzer:2021pfe}~%F.S.G.: new reference
whose goal is to act as a central platform where a large set of information coming from different experiments is gathered. This can be accessed via web or smartphone interface. The~data will be immediately available and will contain relevant information such as the visibility of the event for a given observatory, the~false alarm rate, or~the probability of the event to be of astrophysical~origin.

%%%%%%%%%%%%%%%%%%%%%%%%%%%%%
%%%%%%%%%%%%%%%%%%%%%%%%%%%%%

In addition to these prompt alert systems across collaborations, there are alert follow-up programs like TaToO~\cite{ANTARES:2015fce}, where the most promising ANTARES events trigger a prompt optical, radio, and X-ray follow-up on the sky region where the neutrino candidate comes from, looking for any potential transient counterpart within hours, days, and months since the event. This is done using a network of optical telescopes at different locations (at a rate of 25 alerts per year) and, for~the most energetic ones (around 6 alerts per year), the~XRT instrument aboard the Swift satellite and the Murchison Wide field Array radio telescope~\cite{tatoo:2021}.

As another example of how data from different experiments is shared, we have the Astrophysical Multimessenger Observatory Network (AMON)~\cite{AyalaSolares:2019iiy}. One of the main ideas of AMON is to use sub-threshold data from different experiments to exploit the fact that a combined detection is expected to increase the significance of the event. Therefore, an~event that by itself is not enough to claim a detection with a single experiment, and~could have been rejected, can actually become significant when detected in coincidence with other observatories. An~example of a recent analysis done through this network is~\cite{Hugo:2021}, focused on gamma-ray and neutrino~coincidences.

\subsection{Future~Challenges}
Concerning the multimessenger astronomy goals in the next few years, one thing that should be attainable very soon is a firm confirmation of a source of cosmic neutrinos above the discovery threshold of 5$\sigma$. To~this end, the~selection of the most promising sources, to~reduce the amount of trials in the search, thanks to multimessenger observations will be crucial.
This is quite likely to be accomplished by more than one project which will provide an unbiased way of measuring the spectrum of the sources, bringing key information to understand the high-energy neutrino production and acceleration mechanisms in the source. 
Also combined analyses are possible, as~has been already done in the past~\cite{ANTARES:2020srt}.

\begin{figure}[H]
\vspace{-9pt}
\includegraphics[width=\linewidth]{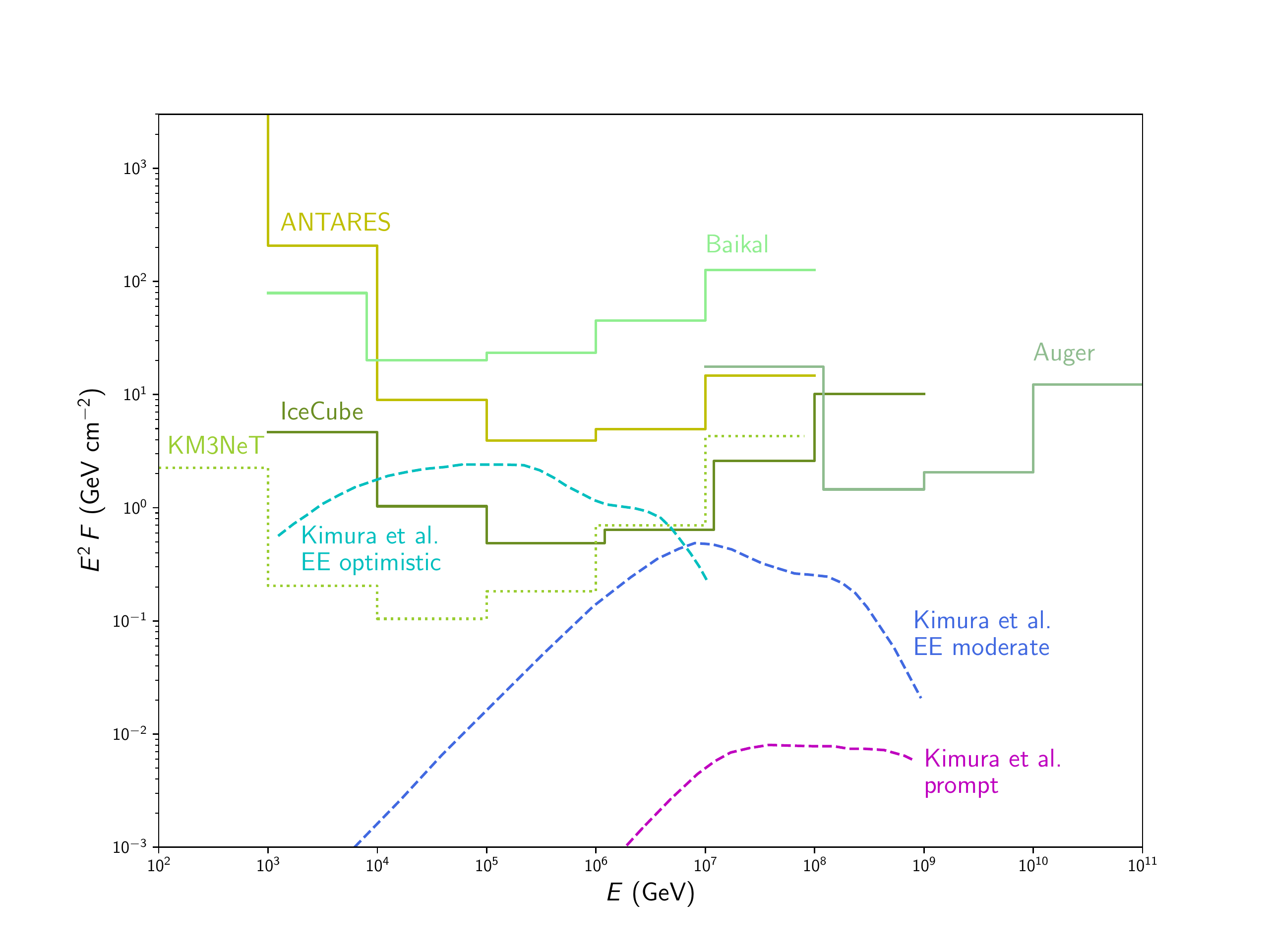}
\caption{Fluence upper limits, per flavor, on~the high-energy neutrino emission from experimental data assuming a $\pm$500s time window around the GW170817 {event}
~\cite{ANTARES:2017bia}. Baikal limits are from~\cite{Baikal:2019}. KM3NeT preliminary sensitivity, computed for an optimal zenith angle, is from~\cite{Palacios:2021}. Theoretical models for comparison are from~\cite{Kimura:2018vvz}. Figure adapted from the work in ~\cite{ANTARES:2017bia}.}
\label{fig2}
\end{figure}

One multimessenger observation that is most awaited, and~can probably be achieved thanks to the improved sensitivity of the future experiments, is the coincident detection of a GW event with high-energy neutrinos. 
Thanks to recent GW observations, we have a better understanding of the link between neutron star mergers and short GRBs, and~the physics involved. We already discussed the particular case of the binary neutron star merger GW170817, which led to the GRB170817A coincidence. This type of event should produce a GW signature together with gamma rays and neutrinos. However, at~present, the~only confirmed coincident detection is the GW-gamma correlation, while no evidence of neutrino emission was found~\cite{ANTARES:2017bia,Baikal:2019}.
The theoretical estimations of neutrino production~\cite{Kimura:2018vvz} from GW170817 show that the expected flux should be already detectable, with~current neutrino telescopes, under~favorable circumstances, see Figure~\ref{fig2}.

%%%%%%%%%%%%%%%%%%%%%%%%%%%%%
%%%%%%%%%%%%%%%%%%%%%%%%%%%%%

Apart from the detection of high-energy neutrinos, the~detection of the prompt emission in gamma-rays by observatories like HAWC or LHAASO will be essential to understand how these energetic explosions~work.

\begin{figure}[H]
\includegraphics[width=0.97\linewidth]{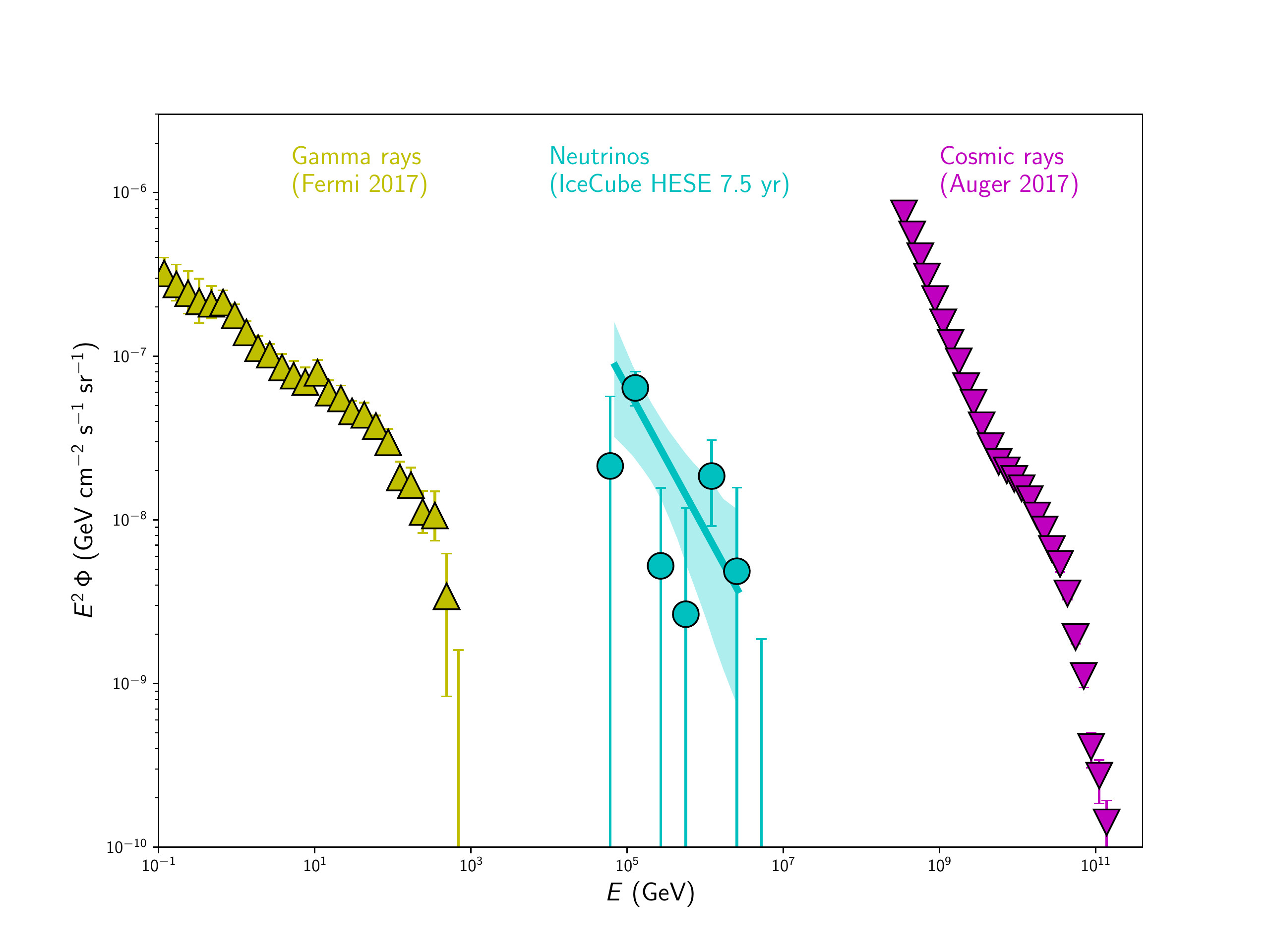}
\caption{High-energy fluxes of gamma {rays}~\cite{Fermi-LAT:2014ryh}, neutrinos~\cite{IceCube:2020wum}, and~cosmic rays~\cite{Auger:2017}. Figure adapted from~the work in \cite{IceCube:2020wum}.}
\label{fig3}
\end{figure}

Another open question to be addressed by multimessenger observations is the connection between ultra-high-energy CRs, gamma-rays, and~high-energy neutrinos. 
Intensity of gamma-rays, neutrinos, and UHECRs has been shown to be comparable (see \mbox{Figure~\ref{fig3}}), suggesting that they may be powered by the same sources. 
As blazars are the most abundant extragalactic sources, they are by default the most promising candidates. However, blazars do not seem to fit the bill, as they are subdominant in the high-energy neutrino flux~\cite{Oikonomou:2021}.

Apart from the usual suspects (e.g., GRBs and TDEs), the~case for star-forming galaxies as common sources has recently grown in popularity thanks to the work in~\cite{Roth:2021lvk} where the authors claim that the diffuse gamma-ray flux detected by Fermi-LAT~\cite{Fermi-LAT:2014ryh} is actually dominated by star-forming galaxies.
At the same time, the~data collected in the Pierre Auger Observatory showed an indication of anisotropy at 4.0$\sigma$ level in the arrival direction of CRs with E > 39 EeV showing an excess from the direction of nearby starburst galaxies~\cite{PierreAuger:2018qvk}. Claiming that this type of high-rate star-forming galaxies can be the cause of $\sim$~10\% of the ultra-high-energy CR flux.
It has been also shown that the contribution of the starburst galaxies to the neutrino diffuse flux is sub-dominant and constrained to be at the level of $\sim$10\%~\cite{Lunardini:2019zcf}. Therefore, it remains unanswered for now whether there is a dominant type of source accelerating all cosmic~messengers.

%%%%%%%%%%%%%%%%%%%%%%%%%%%%%

%%%%%%%%%%%%%%%%%%%%%%%%%%%%%

Finally, if~we restrict the search to our Galaxy, one of the questions that will be solved in the near future thanks to multimessenger observations is what are the Galactic CR sources and how those CR propagate through the Galaxy.
Regarding this, it has been shown that a sub-PeV diffuse Galactic gamma-ray emission exists~\cite{TibetASgamma:2021tpz}.
Based on this result in~\cite{Fang:2021ylv}, the authors showed that the Galactic neutrino contribution should constitute roughly 5--10$\%$ of the IceCube diffuse flux and that, in~the 10--100 TeV range, the~expected Galactic neutrino flux should be comparable to the total neutrino diffuse flux. If~so, the~next-generation neutrino telescopes should be sensitive enough to detect it.
It is possible that part of the measured gamma-ray diffuse flux comes from individual sources, being the obvious candidates the very-high-energy sources detected by HAWC~\cite{HAWC:2019tcx} and LHAASO~\cite{Cao:2021}. 
Combined multimessenger observations should be able to confirm whether this is actually the case.
In fact, there have been claims for evidence for such a joint production, of~high-energy neutrinos and gamma rays, in~the Cygnus Coocon region, based on the correlation of a high-energy IceCube neutrino and a high-energy (>300 TeV) photon flare observed by the Carpet–2 experiment~\cite{Carpet-3Group:2021ygp}. 

%%%%%%%%%%%%%%%%%%%%%%%%%%%%%%%%%%%%%%%%%%
\section{Outlook}
Multimessenger astronomy is a new branch of astroparticle physics, for~which neutrinos are expected to play a key role. It took just a few events, detected in coincidence, to~demonstrate that combining different messengers has a great potential for~discoveries.

Questions like what is the origin of the ultra-high energy cosmic rays are likely to be answered thanks to multimessenger~observations.

Considering the numerous facilities planned for the near future, or~already taking data (KM3NeT, IceCube-Gen2, GVD-Baikal, P-ONE, CTA, LHAASO, KAGRA, AugerPrime, etc.), multimessenger astronomy has a bright future and is expected to revolutionize our understanding of the Universe in the next~decade.

%%%%%%%%%%%%%%%%%%%%%%%%%%%%%%%%%%%%%%%%%%
\vspace{6pt} 
\authorcontributions{{All authors contributed equally to this article. All authors have read and agreed to the published version of the manuscript.}}%MDPI:please confirm if these parts should be added. For research articles with several authors, a short paragraph specifying their individual contributions must be provided. The following statements should be used ``Conceptualization, X.X. and Y.Y.; methodology, X.X.; software, X.X.; validation, X.X., Y.Y. and Z.Z.; formal analysis, X.X.; investigation, X.X.; resources, X.X.; data curation, X.X.; writing---original draft preparation, X.X.; writing---review and editing, X.X.; visualization, X.X.; supervision, X.X.; project administration, X.X.; funding acquisition, Y.Y. All authors have read and agreed to the published version of the manuscript.'', please turn to the  \href{http://img.mdpi.org/data/contributor-role-instruction.pdf}{CRediT taxonomy} for the term explanation. Authorship must be limited to those who have contributed substantially to the work~reported.

\funding{This work was funded by Generalitat Valenciana, Spain (CIDEGENT/2018/034 and CIDEGENT/2020/049 grants).}

\institutionalreview{{Not applicable.}}%In this section, please add the Institutional Review Board Statement and approval number for studies involving humans or animals. Please note that the Editorial Office might ask you for further information. Please add ``The study was conducted according to the guidelines of the Declaration of Helsinki, and approved by the Institutional Review Board (or Ethics Committee) of NAME OF INSTITUTE (protocol code XXX and date of approval).'' OR ``Ethical review and approval were waived for this study, due to REASON (please provide a detailed justification).'' OR ``Not applicable'' for studies not involving humans or animals. You might also choose to exclude this statement if the study did not involve humans or animals.

\informedconsent{{Not applicable.}}%Any research article describing a study involving humans should contain this statement. Please add ``Informed consent was obtained from all subjects involved in the study.'' OR ``Patient consent was waived due to REASON (please provide a detailed justification).'' OR ``Not applicable'' for studies not involving humans. You might also choose to exclude this statement if the study did not involve humans.Written informed consent for publication must be obtained from participating patients who can be identified (including by the patients themselves). Please state ``Written informed consent has been obtained from the patient(s) to publish this paper'' if applicable.

\dataavailability{{Not applicable.}} %In this section, please provide details regarding where data supporting reported results can be found, including links to publicly archived datasets analyzed or generated during the study. Please refer to suggested Data Availability Statements in section ``MDPI Research Data Policies'' at \url{https://www.mdpi.com/ethics}. You might choose to exclude this statement if the study did not report any data.

\acknowledgments{The authors thank the Valencia Experimental Group on Astroparticle Physics (VEGA) and the Ministerio de Ciencia, Innovación, Investigación y Universidades (MCIU): Programa Estatal de Generación de Conocimiento (ref. PGC2018-096663-B-C41) (MCIU/FEDER). }

\conflictsofinterest{The authors declare no conflict of interest.} %\conflictsofinterest{Declare conflicts of interest or state ``The authors declare no conflict of interest.'' Authors must identify and declare any personal circumstances or interest that may be perceived as inappropriately influencing the representation or interpretation of reported research results. Any role of the funders in the design of the study; in the collection, analyses or interpretation of data; in the writing of the manuscript, or in the decision to publish the results must be declared in this section. If there is no role, please state ``The funders had no role in the design of the study; in the collection, analyses, or interpretation of data; in the writing of the manuscript, or in the decision to publish the~results''.} 

%%%%%%%%%%%%%%%%%%%%%%%%%%%%%%%%%%%%%%%%%%
%% Only for journal Encyclopedia
%\entrylink{The Link to this entry published on the encyclopedia platform.}

%%%%%%%%%%%%%%%%%%%%%%%%%%%%%%%%%%%%%%%%%%
%% Optional
\abbreviations{Abbreviations}{
The following abbreviations are used in this manuscript:\\

\noindent 
\begin{tabular}{@{}ll}
GW & Gravitational Wave\\
CR & Cosmic Ray\\
GRB & Gamma-Ray Burst\\
TDE & Tidal Disruption Event\\
PMT & Photomultipliers tubes

\end{tabular}}

%%%%%%%%%%%%%%%%%%%%%%%%%%%%%%%%%%%%%%%%%%
%% Optional
\appendixtitles{no} % Leave argument "no" if all appendix headings stay EMPTY (then no dot is printed after "Appendix A"). If~the appendix sections contain a heading then change the argument to~"yes".

%% If you want to add notes in main text, please insert \endnote{} and release below codes.
 %Output endnote listing
\begin{adjustwidth}{-4.6cm}{0cm}
\printendnotes[custom]
\end{adjustwidth}
%%%%%%%%%%%%%%%%%%%%%%%%%%%%%%%%%%%%%%%%%%

%%%%%%%%%%%%%%%%%%%%%%%%%%%%%%%%%%%%%%%%%%
\end{paracol} %do not delete this, it ends the \begin{paracol} in the class file
\reftitle{References}


%\externalbibliography{yes}
%\bibliography{your_external_BibTeX_file}

\begin{thebibliography}{999}

%Ref 1
\bibitem[PDG(2020)]{ParticleDataGroup:2020ssz} Zyla, P.A. et~al. [Particle Data Group]. %MDPI:We modified the format of highlighted references. Please confirm.
\textls[-25]{ Review of Particle Physics. {\em Prog. Theor. Exp. Phys.} {\bf 2020}, {\em 8}, 083C01.% {https://journals.aps.org/prd/abstract/10.1103/PhysRevD.98.030001}.
}

%Ref 2
\bibitem[Davis(1968)]{Davis:1968cp} Davis, R., Jr.; Harmer, D.S.; Hoffman, K.C. Search for neutrinos from the sun. {\em Phys. Rev. Lett.} {\bf 1968}, {\em 20}, 1205, {https://doi.org/10.1103/PhysRevLett.20.1205}.

%Ref 3
\bibitem[SK(1988)]{Hirata:1988ad}  Hirata, K.S.; Kajita, T.; Koshiba, M.; Nakahata, M.; Oyama, Y.; Sato, N.; Suzuki, A.; Takita, M.; Totsuka, Y.; Kifune, T.; et~al. Observation in the Kamiokande-II Detector of the Neutrino Burst from Supernova SN 1987a. {\em Phys. Rev. D} {\bf 1988}, {\em 38}, 448, {https://doi.org/10.1103/PhysRevD.38.448}.

%Ref 4
\bibitem[IMB(1987)]{IMB:1987klg} Haines, T. et~al. [IMB Collaboration]. Neutrinos from SN1987a in the IMB detector. {\em Nucl. Instrum. Meth. A} {\bf 1988}, {\em 264}, 28--31, {https://doi.org/10.1016/0168-9002(88)91097-2}.

%Ref 5
%\bibitem[Baksan(1987)]{Baksan:1987} Alekseev, E.~et~al. Detection of the Neutrino Signal from Supernova 1987A Using the INR Baksan Underground Scintillation Telescope. Proceedings of the ESO Workshop on the SN 1987 A, Munich, Germany, July 6-8, 1987. Editors, I.J. Danziger; Publisher, European Southern Observatory, Garching bei Munchen, Federal Republic of Germany, 1987
\bibitem[Baksan(1987)]{Baksan:1987} Alekseev, E.; Alexeyeva, L.N.; Krivosheina, I.V.; Volchenko, V.I. Detection of the neutrino signal from SN 1987A in the LMC using the INR Baksan underground scintillation telescope. {\em Phys. Lett. B} {\bf 1988}, {\em 205}, 209--214, {https://doi.org/10.1016/0370-2693(88)91651-6}.

%Ref 6
\bibitem[DUMAND(1989)]{DUMAND:1989dxw} Babson, J. et al. [DUMAND Collaboration]. Cosmic Ray Muons in the Deep Ocean. {\em Phys. Rev. D} {\bf 1990}, {\em 42}, 3613, {https://doi.org/10.1103/PhysRevD.42.3613}.

%Ref 7
\bibitem[IceCube(2014)]{IceCube:2014stg} Aartsen, M.G. et al. [IceCube Collaboration]. Observation of High-Energy Astrophysical Neutrinos in Three Years of IceCube Data. {\em Phys. Rev. Lett.} {\bf 2014}, {\em 113}, 101101, {https://doi.org/10.1103/PhysRevLett.113.101101}.

%Ref 8
\bibitem[LIGO(2016)]{LIGOScientific:2016aoc} Abbott, B.P. et al. [LIGO Scientific Collaboration and Virgo Collaboration]. Observation of Gravitational Waves from a Binary Black Hole Merger. {\em Phys. Rev. Lett.} {\bf 2016}, {\em 116}, 061102, {https://doi.org/10.1103/PhysRevLett.116.061102}.

%Ref 9
\bibitem[LIGO(2017)]{LIGOScientific:2017vwq} Abbott, B.P. et al. LIGO Scientific Collaboration and Virgo Collaboration]. GW170817: Observation of Gravitational Waves from a Binary Neutron Star Inspiral. {\em Phys. Rev. Lett.} {\bf 2017}, {\em 119}, 161101, {https://doi.org/10.1103/PhysRevLett.119.161101}.

%Ref 10
\bibitem[LIGOmm(2017)]{LIGOScientific:2017ync} Abbott, B.P.  et~al. [AST3, CAASTRO, VINROUGE, MASTER, J-GEM, GROWTH, JAGWAR, CaltechNRAO, TTU-NRAO, NuSTAR, Pan-STARRS, MAXI Team, TZAC Consortium, KU, Nordic Optical Telescope, ePESSTO, GROND, Texas Tech University, SALT Group, TOROS, BOOTES, MWA, CALET, IKI-GW Follow-up, H.E.S.S., LOFAR, LWA, HAWC, Pierre Auger, ALMA, Euro VLBI Team, Pi of Sky, Chandra Team at McGill University, DFN, ATLAS Telescopes, High Time Resolution Universe Survey, RIMAS, RATIR, SKA South Africa/MeerKAT].  Multi-messenger Observations of a Binary Neutron Star Merger. {\em Astrophys. J. Lett.} {\bf 2017}, {\em 848}, L12, {https://doi.org/10.3847/2041-8213/aa91c9}.

%Reference 11
\bibitem[Metzger(2017)]{Metzger:2017} Metzger, B.D. Kilonovae. {\em Living Rev. Relativ.} {\bf 2017}, {\em 20}, 3, {https://doi.org/10.1007/s41114-017-0006-z}.

% Reference 12
\bibitem[LiXin(1998)]{Lixin:1998} Li-Xin, L.; Paczy\'{n}ski, B. \textls[-15]{Transient Events from Neutron Star Mergers. {\em Astrophys. J.} {\bf 1998}, {\em 507}, L59--L62, {https://doi.org/10.1086/311680}.}

%Ref 13
\bibitem[IceCubemm(2017)]{IceCube:2018dnn} Aartsen, M.G. et al. [IceCube Collaboration]. Multimessenger observations of a flaring blazar coincident with high-energy neutrino IceCube-170922A. {\em Science} {\bf 2018}, {\em 361}, eaat1378, 
{https://doi.org/10.1126/science.aat1378}.

%Ref 14
\bibitem[IceCube(2017)]{IceCube:2018cha} Aartsen, M.G. et al. [IceCube Collaboration]. Neutrino emission from the direction of the blazar TXS 0506+056 prior to the IceCube-170922A alert. {\em Science} {\bf 2018}, {\em 361}, 147--151,  	{https://doi.org/10.1126/science.aat2890}.

%Reference 15
\bibitem[Satalecka(2021)]{Satalecka:2021} Satalecka, K.~et~al. [MAGIC, ATCA, OVRO and TELAMON Collaborations]. Multi-epoch monitoring of TXS 0506+056 with MAGIC and
MWL partners. In { Proceedings of the International Cosmic Ray Conference (ICRC2021), Berlin, Germany, 12--23 July 2021};{Volume 395}, p. 875.
% \url{https://pos.sissa.it/395/875/}.

%Ref 16
\bibitem[Illuminati(2021b)]{Illuminati:2021b} Illuminati, G. ~et~al. [ANTARES Collaboration]. Searches for point-like sources of cosmic neutrinos with 13 years of ANTARES data. In Proceedings of the International Cosmic Ray Conference (ICRC2021), Berlin, Germany, 12--23 July 2021; Volume 395, p.1161.
%\url{https://pos.sissa.it/395/1161}.

%Ref 17
\bibitem[IceCube(2019)]{IceCube:2019cia} Aartsen, M.G.; ~Ackermann, M.; Adams, J.; Aguilar, J.A.; Ahlers, M.; Ahrens, M.; Alispach, C.; Andeen, K.; Anderson, T.; Ansseau, I.; et~al. Time-Integrated Neutrino Source Searches with 10 Years of IceCube Data. {\em Phys. Rev. Lett.} {\bf 2020}, {\em 124}, 051103, {https://doi.org/10.1103/PhysRevLett.124.051103}.

%Ref 18
\bibitem[Keivani(2018)]{Keivani:2018rnh} Keivani, A.; ~Murase, K.; Petropoulou, M.; Fox, D.B.; Cenko, S.B.; Chaty, S.; Coleiro, A.; DeLaunay, J.J.; Dimitrakoudis, S.; Evans, P.A.; et~al. A Multimessenger Picture of the Flaring Blazar TXS 0506+056: Implications for High-Energy Neutrino Emission and Cosmic Ray Acceleration. {\em Astrophys. J.} {\bf 2018}, {\em 864}, 84, {https://doi.org/10.3847/1538-4357/aad59a}.

% Reference 19
\bibitem[Xue(2019)]{Xue:2019txw} Xue, R.; Liu, R.Y.; Petropoulou, M.; Oikonomou, F.; Wang, Z.R.; Wang, K.; Wang, X.Y. A Two-zone Model for Blazar Emission: Implications for TXS 0506+056 and the Neutrino Event IceCube-170922A. {\em Astrophys. J.} {\bf 2019}, {\em 886}, 23, {https://doi.org/10.3847/1538-4357/ab4b44}.

% Reference 20
\bibitem[Plavin(2020b)]{Plavin:2020mkf} Plavin, A.; Kovalev, Y.Y.; Kovalev, Y.A.; Troitsky, S.V. Directional Association of TeV to PeV Astrophysical Neutrinos with Radio Blazars. {\em Astrophys. J.} {\bf 2021}, {\em 908}, 157, {https://doi.org/10.3847/1538-4357/abceb8}.

% Reference 21
\bibitem[Plavin(2020a)]{Plavin:2020emb} Plavin, A.;~Plavin, A.; Kovalev, Y.Y.; Kovalev, Y.A.; Troitsky, S. Observational Evidence for the Origin of High-energy Neutrinos in Parsec-scale Nuclei of Radio-bright Active Galaxies. {\em Astrophys. J.} {\bf 2020}, {\em 894}, 101, {https://doi.org/10.3847/1538-4357/ab86bd}.

%Ref 22
\bibitem[ANTARES(2020)]{ANTARES:2020zng} Albert, A.; ~Andr\'{e}, M.; Anghinolfi, M.; Anton, G.; Ardid, M.; Aubert, J.J.; Aublin, J.; Baret, B.; Basa, S.; Belhorma, B.; et~al. ANTARES Search for Point Sources of Neutrinos Using Astrophysical Catalogs: A Likelihood Analysis. {\em Astrophys. J.} {\bf 2021}, {\em 911}, 48, {https://doi.org/10.3847/1538-4357/abe53c}.

%Reference 23
\bibitem[Illuminati(2021)]{Illuminati:2021} Illuminati, G. ~et~al. [ANTARES Collaboration]. ANTARES search for neutrino flares from the direction of radio-bright blazars. In Proceedings of the International Cosmic Ray Conference (ICRC2021), Berlin, Germany, 12--23 July 2021; {Volume 395}, p. 972.
%\url{https://pos.sissa.it/395/972}

% Reference 24
\bibitem[IceCube(2016)]{IceCube:2016cqr} Aartsen, M.G.;~Ackermann, M.; Adams, J.; Aguilar, J.A.; Ahlers, M.; Ahrens, M.; Altmann, D.; Andeen, K.; Anderson, T.; Ansseau, I.; et~al. The IceCube Realtime Alert System. {\em Astropart. Phys.} {\bf 2017}, {\em 92}, 30, {https://doi.org/10.1016/\linebreak j.astropartphys.2017.05.002}.

%Ref 25
\bibitem[Rodrigues(2020)]{Rodrigues:2020fbu} Rodrigues, X.;~Garrappa, S.; Gao, S.; Paliya, V.S.; Franckowiak, A.; Winter, W. Multiwavelength and Neutrino Emission from Blazar PKS 1502 + 106. {\em Astrophys. J.} {\bf 2021}, {\em 912}, 54, {https://doi.org/10.3847/1538-4357/abe87b}.

%Ref 26
\bibitem[Paliya(2020)]{Paliya:2020mqm} Paliya, V.S.; ~B\"{o}ttcher, M.; Olmo-Garc\'{i}a, A.; Dom\'{i}nguez, A.; de Paz, A.G.; Franckowiak, A.; Garrappa, S.; Stein, R. Multifrequency Observations of the Candidate Neutrino-emitting Blazar BZB J0955+3551. {\em Astrophys. J.} {\bf 2020}, {\em 902}, 29, {https://doi.org/10.3847/1538-4357/abb46e}.

%Ref 27
\bibitem[Giommi(2020)]{Giommi:2020viy} Giommi, P.;~Padovani, P.; Oikonomou, F.; Glauch, T.; Paiano, S.; Resconi, E. 3HSP J095507.9+355101: A flaring extreme blazar coincident in space and time with IceCube-200107A.
{\em Astron. Astrophys.} {\bf 2020}, {\em 640}, L4, {https://doi.org/ 	10.1051/0004-6361/202038423}.

%Ref 28
\bibitem[Fermi-LAT(2019)]{Fermi-LAT:2019pir} Ajello, M.;~Angioni, R.; Axelsson, M.; Ballet, J.; Barbiellini, G.; Bastieri, D.; Gonzalez, J.B.; Bellazzini, R.; Bissaldi, E.; Bloom, E.D.; et al. The Fourth Catalog of Active Galactic Nuclei Detected by the Fermi Large Area Telescope. {\em Astrophys. J.} {\bf 2020}, {\em 892}, 105, {https://doi.org/10.3847/1538-4357/ab791e}.

%Ref 29
\bibitem[IceCube(2016)]{IceCube:2016qvd} Aartsen, M.G. ~et~al. [IceCube Collaboration]. The contribution of Fermi-2LAC blazars to the diffuse TeV-PeV neutrino flux. {\em Astrophys. J.} {\bf 2017}, {\em 835}, 45, {https://doi.org/10.3847/1538-4357/835/1/45}.

% Reference 30
\bibitem[Wang(2011)]{Wang:2011} Wang, X.Y.;~Liu, R.Y.; Dai, Z.G.; Cheng, K.S. Probing the tidal disruption flares of massive black holes with high-energy neutrinos. {\em Phys. Rev. D} {\bf 2011}, {\em 84}, 081301(R), {https://doi.org/10.1103/PhysRevD.84.081301}.

% Reference 31
\bibitem[Bellm(2019)]{Bellm:2019} Bell, E.C.;~Kulkarni, S.R.; Graham, M.J.; Dekany, R.; Smith, R.M.; Riddle, R.; Masci, F.J.; Helou, G.; Prince, T.A.; Adams, S.M.; et~al. The Zwicky Transient Facility: System Overview, Performance, and First Results. {\em Publ. Astron. Soc. Pac.} {\bf 2019}, {\em 131}, 018002, {https://doi.org/10.1088/1538-3873/aaecbe}.

% Reference 32
\bibitem[Stein(2020)]{Stein:2020xhk} Stein, R.;~van Velzen, S.; Kowalski, M.; Franckowiak, A.; Gezari, S.; Miller-Jones, J.C.; Frederick, S.; Sfaradi, I.; Bietenholz, M.F.; Horesh, A.; et~al. A tidal disruption event coincident with a high-energy neutrino. {\em Nat. Astron.} {\bf 2021}, {\em 908}, 510--518, {https://doi.org/10.1038/s41550-020-01295-8}.

% Reference 33 %NEW reference!
\bibitem[ANTARES(2021)]{ANTARES:2021jmp} Albert, A. et al. [ANTARES Collaboration]. Search for Neutrinos from the Tidal Disruption Events AT2019dsg and AT2019fdr with the ANTARES Telescope. {\em Astrophys. J.} {\bf 2021}, {\em 920}, 50, {https://doi.org/10.3847/1538-4357/ac16d6}. 

%Reference 34
\bibitem[Allakhverdyan(2021)]{Allakhverdyan:2021} Allakhverdyan, V.A.;~et~al. [GVD-Baikal Collaboration]. Multi-messenger and real-time astrophysics with the Baikal-GVD telescope. In { Proceedings of the International Cosmic Ray Conference (ICRC2021), Berlin, Germany, 12--23 July 2021}; {Volume 395}, p. 946. 
%\url{https://pos.sissa.it/395/946}.

%Reference 35
\bibitem[Stein(2021)]{Stein:2021} Stein, R. Tidal disruption event coincident with a high-energy neutrino. In { Proceedings of the International Cosmic Ray Conference (ICRC2021), Berlin, Germany, 12--23 July 2021}; %Publisher: \hl{Italy}, {2021}; 
{Volume 395}, p. 009. 
%\url{https://pos.sissa.it/395/009}.

%Ref 36
\bibitem[Waxman(1997)]{Waxman:1997ti} Waxman, E.; Bahcall, J.N. High-energy neutrinos from cosmological gamma-ray burst fireballs. {\em Phys. Rev. Lett.} {\bf 1997}, {\em 78}, 2292, {https://doi.org/10.1103/PhysRevLett.78.2292}.

%Ref 37
\bibitem[ANTARES(2017)]{Albert:2016eyr} Albert, A.; Andr\'{e}, M.; Anghinolfi, M.; Anton, G.; Ardid, M.; Aubert, J.J.; Avgitas, T.; Baret, B.; Barrios-Mart\'{i}, J.; Basa, S.; et~al. Search for high-energy neutrinos from bright GRBs with ANTARES. {\em Mon. Not. Roy. Astron. Soc.} {\bf 2017}, {\em 469}, 906--915, {https://doi.org/10.1093/mnras/stx902}.

%Ref 38
\bibitem[IceCube(2016)]{IceCube:2016ipa} Aartsen, M.G. et al. [IceCube Collaboration]. An All-Sky Search for Three Flavors of Neutrinos from Gamma-Ray Bursts with the IceCube Neutrino Observatory. {\em Astrophys. J.} {\bf 2016}, {\em 824}, 2, 115, {https://doi.org/10.3847/0004-637X/824/2/115}.

%Ref 39
\bibitem[Bartos(2021)]{Bartos:2021tok} Bartos, I.; Veske, D.; Kowalski, M.; Marka, Z.; Marka, S.  The IceCube Pie Chart: Relative Source Contributions to the Cosmic Neutrino Flux. {\em arXiv} {\bf 2021}, arXiv:2105.03792. %(accessed on 30-09-2021).
%Available online: \url{https://arxiv.org/abs/2105.03792v1} ().

% Reference 40
\bibitem[KM3NeT(2016)]{KM3Net:2016zxf} Adrian-Martinez, S.;~Ageron, M.; Aharonian, F.; Aiello, S.; Albert, A.; Ameli, F.; Anassontzis, E.; Andre, M.; Androulakis, G.; Anghinolfi, M.;  et~al. Letter of intent for KM3NeT 2.0. {\em J. Phys. G.} {\bf 2016}, {\em 43}, 084001, {https://doi.org/10.1088/0954-3899/43/8/084001}.

%Reference 41
\bibitem[Baikal(2021)]{Baikal:2021} Belolaptikov, I.;~Dzhilkibaev, Z. [GVD-Baikal Collaboration]. Neutrino Telescope in Lake Baikal: Present and Nearest Future. In Proceedings of the International Cosmic Ray Conference (ICRC2021), Berlin, Germany, 12--23 July 2021; %Publisher:  \hl{Italy}, { 2021}, 
{Volume 395}, p. 002.
%\url{https://pos.sissa.it/395/002}.

%Reference 42
\bibitem[IceCube(2021)]{IceCube:2021} Kowalski, M.;~et~al. [IceCube Collaboration]. IceCube: The Window to the Extreme Universe. In { Proceedings of the International Cosmic Ray Conference (ICRC2021), Berlin, Germany, 12--23 July 2021}; %Publisher: \hl{Italy}, {2021}; 
{Volume 395}, p. 022.
%\url{https://pos.sissa.it/395/022}.

% Reference 43
\bibitem[IceCube-Gen2(2020)]{IceCube-Gen2:2020qha} Aartsen, M.G.;~Abbasi, R.; Ackermann, M.; Adams, J.; Aguilar, J.A.; Ahlers, M.; Ahrens, M.; Alispach, C.; Allison, P.; Amin, N.M.; et~al. IceCube-Gen2: The window to the extreme Universe {\em J. Phys. G.} {\bf 2021}, {\em 48}, 060501, {https://doi.org/10.1088/\linebreak 1361-6471/abbd48}.

% Reference 44
\bibitem[IceCube(2019)]{IceCube:2019xdf} Aartsen, M.G.~et~al. [IceCube Collaboration]. The IceCube Neutrino Observatory---Contributions to the 36th International Cosmic Ray Conference (ICRC2019). {\em arXiv} {\bf 2019}, arXiv:1907.11699. %(accessed on 30-09-2021).
%Available online: \url{https://arxiv.org/abs/1907.11699v1} (\hl{accessed on}). %MDPI:Please add accessed date.

% Reference 45
\bibitem[IceCube-Gen2(2016)]{IceCube:2016xxt} Aartsen, M.G.;~Abraham, K.; Ackermann, M.; Adams, J.; Aguilar, J.A.; Ahlers, M.; Ahrens, M.; Altmann, D.; Andeen, K.; Anderson, T.; et~al. PINGU: A Vision for Neutrino and Particle Physics at the South Pole {\em J. Phys. G.} {\bf 2017}, {\em 44}, 054006, {https://doi.org/10.1088/1361-6471/44/5/054006}.

%Reference 46
\bibitem[PONE(2021)]{PONE:2021} Resconi, E. The Pacific Ocean Neutrino Experiment at Ocean Networks Canada. In {Proceedings of the International Cosmic Ray Conference (ICRC2021), Berlin, Germany, 12--23 July 2021}; %Publisher: \hl{Italy}, {2021};
 {Volume 395}, p. 024.
%\url{https://pos.sissa.it/395/024}.

%Ref 47
\bibitem[Hyper-Kamiokande(2018)]{Hyper-Kamiokande:2018ofw} Abe, K.;~et~al. [Hyper-Kamiokande Proto-Collaboration]. Hyper-Kamiokande Design Report. {\em arXiv} {\bf 2018}, arXiv:1805.04163. %(accessed on 30-09-2021).
%Available online: \url{https://arxiv.org/abs/1805.04163v2} (\hl{accessed on}). %MDPI:Please add accessed date.

%Ref 48
\bibitem[DUNE(2020)]{DUNE:2020lwj} Abi, B.;~Acciarri, R.;  Acero, M.A.; Adamov, G.;  Adams, D.;  Adinolfi, M.;  Ahmad, Z.; Ahmed,  J.; Alion, T.;  Alonso Monsalve, S.; et~al. Deep Underground Neutrino Experiment (DUNE), Far Detector Technical Design Report, Volume I Introduction to DUNE.  {\em J. Instrum.} {\bf 2020}, {\em 15}, T08008, {https://doi.org/10.1088/1748-0221/15/08/T08008}.

%Ref 49
\bibitem[CTA(2021)]{CTA:2021} Zanin, R.;~Abdalla, H.; Abe, H.; Abe, S.; Abusleme, A.; Acero, F.; Acharyya, A.; Acin Portella, V.;  Ackley, K.; Adam, R.; et~al. CTA---the World’s largest ground-based gamma-ray observatory. In {Proceedings of the International Cosmic Ray Conference (ICRC2021)}, Berlin, Germany, 12--23 July 2021; %Publisher: \hl{Italy}, {2021}; 
{Volume 395}, p. 005.
%\url{https://pos.sissa.it/395/005}.

%Ref 50
\bibitem[LHAASO(2021)]{LHAASO:2021} Cao, Z. Highlights of LHAASO science results. In {Proceedings of the International Cosmic Ray Conference (ICRC2021), Berlin, Germany, 12--23 July 2021}; %Publisher: \hl{Italy}, 2021; 
{Volume 395}, p. 011.
%\url{https://pos.sissa.it/395/011}.

%Ref 51
\bibitem[KAGRA(2019)]{KAGRA:2019} Akutsu, T.;~et~al. [KAGRA collaboration].  KAGRA: 2.5 generation interferometric gravitational wave detector. {\em Nat. Astron.} {\bf 2019}, {\em 3}, 35--40, {https://doi.org/10.1038/s41550-018-0658-y}.

%Ref 52
\bibitem[PierreAuger(2016)]{PierreAuger:2016qzd} Aab, A.;~et~al. [The Pierre Auger Collaboration]. The Pierre Auger Observatory Upgrade---Preliminary Design Report. {\em arXiv} {\bf 2016}, arXiv:1604.03637. %(accessed on 30 September 2021).

%Available online: \url{https://arxiv.org/abs/1604.03637v1} (\hl{accessed on}). %MDPI:Please add accessed date.

%Ref 53
\bibitem[GCN(2021)]{GCN:2021}
The Gamma-ray Coordinates Network. Available online: \url{https://gcn.gsfc.nasa.gov/} (accessed on 30 September 2021).%MDPI:Please add accessed date.

%Ref 54
\bibitem[ATEL(2021)]{ATEL:2021}
The Astronomer’s Telegram. Available online: \url{https://www.astronomerstelegram.org/} (accessed on 30 September 2021). %MDPI:Please add accessed date.

%Ref 54
%\bibitem[Colibri(2021)]{Colibri:2021} Schüssler, F.;~\hl{Alkan, A.K.; Lefranc, V.; Reichherzer, P.} Astro-COLIBRI: A new platform for real-time multi-messenger astrophysics. In { Proceedings of the International Cosmic Ray Conference (ICRC2021)}, Berlin, Germany, 12--23 July 2021; %Publisher: \hl{Italy}, {2021}, {Volume 395}, p. 935.
%\url{https://pos.sissa.it/395/935}.

%Ref 55 %NEW reference!
\bibitem[Colibri(2021)]{Reichherzer:2021pfe} Reichherzer, P.; Sch\"ussler, F.; Lefranc, V.; Yusafzai, A.; Alkan, A.K.; Ashkar, H.; Becker Tjus, J. Astro-COLIBRI\textemdash{}The COincidence LIBrary for Real-time Inquiry for Multimessenger Astrophysics. {\em Astrophys. J. Supp.} {\bf 2021}, {\em 256},  5, \url{https://doi.org/10.3847/1538-4365/ac1517}.

%Ref 56
\bibitem[ANTARES(2015)]{ANTARES:2015fce} Adri\'an-Mart\'\i{}nez, S.; Ageron, M.; Albert, A.; Al Samarai, I.; Andr\'{e}, M.; Anton, G.; Ardid, M.; Aubert, J.J.; Baret, B.; Barrios-Marti, J.; et al. [ANTARES, TAROT, ROTSE, Swift, Zadko Collaborations]. Optical and X-ray early follow-up of ANTARES neutrino alerts. {\em J. Cosmol. Astropart. Phys.} {\bf 2016}, {\em 2016}, 062, {https://doi.org/10.1088/1475-7516/2016/02/062}. %Please confirm if this should be deleted.

%Ref 57
\bibitem[tatoo(2019)]{tatoo:2021} Dornic, D.; Ageron, M.; Bertin, V.; Brunner, J.; Coleiro, A.; Sch\"{u}ssler, F.; Turpin, D.; Vallage, B. Ten years of multi-wavelength follow-up observations of ANTARES neutrino alerts. In {Proceedings of the International Cosmic Ray Conference (ICRC2019), Madison, WI, USA, 24 July--1 August 2019}; %Publisher: \hl{Italy}, {2019};
 {Volume 358}, \mbox{p. 871.}
%\url{https://pos.sissa.it/358/871}.

%Ref 58
\bibitem[AyalaSolares(2019)]{AyalaSolares:2019iiy} Solares, H.A.A.; Coutu, S.; Cowen, D.F.; DeLaunay, J.J.; Fox, D.B.; Keivani, A.; Mostaf\'{a}, M.; Murase, K.; Oikonomou, F.; Seglar-Arroyo, M.; et al. The Astrophysical Multimessenger Observatory Network (AMON): Performance and science program. {\em Astropart. Phys.} {\bf 2020}, {\em 144}, 68--76, {https://doi.org/10.1016/j.astropartphys.2019.06.007}.

%Ref 59
\bibitem[Hugo(2021)]{Hugo:2021} Hugo, A. et al. [the AMON Group, the IceCube Collaboration, the HAWC Collaboration, the ANTARES Collaboration]. Multimessenger NuEM Alerts with AMON. In { Proceedings of the International Cosmic Ray Conference (ICRC2021), Berlin, Germany, 12--23 July 2021}; %Publisher: \hl{Italy}, {2021}; 
{Volume 395}, p. 958.
%\url{https://pos.sissa.it/395/958}.

% Reference 60
\bibitem[ANTARES(2020)]{ANTARES:2020srt} Albert, A.~et~al. [IceCube Collaboration]. ANTARES and IceCube Combined Search for Neutrino Point-like and Extended Sources in the Southern Sky {\em Astrophys. J.} {\bf 2020}, {\em 892}, 92, {https://doi.org/10.3847/1538-4357/ab7afb}.

%Ref 61
\bibitem[ANTARES(2017)]{ANTARES:2017bia} Albert, A.; et al. [LIGO Scientific Collaboration and Virgo Collaboration]. Search for High-energy Neutrinos from Binary Neutron Star Merger GW170817 with ANTARES, IceCube, and the Pierre Auger Observatory. {\em Astrophys. J. Lett.} {\bf 2017}, {\em 850}, L35, {https://doi.org/10.3847/2041-8213/aa9aed}.

%Ref 
%\bibitem[Baikal(2018)]{Baikal-GVD:2018cya} Avrorin, A. D. et al (Baikal-GVD Collaboration) Search for High-Energy Neutrinos from GW170817 with the Baikal-GVD Neutrino Telescope. {\em JETP Lett.} {\bf 2018}, {\em 108}, 12, 787, {https://doi.org/10.1134/S0021364018240025}.

%Ref 62
\bibitem[Baikal(2019)]{Baikal:2019} Avrorin, A.D.; Avrorin, A.V.; Aynutdinov, V.M.; Bannash, R.; Belolaptikov, I.A.; Brudanin, V.B.; Budnev, N.M.; Domogatsky, G.V.; Doroshenko, A.A.; Dvornicky, R.; et al. [GVD-Baikal Collaboration]. The Baikal-GVD neutrino telescope: First results of multi-messenger studies---Contributions to the 36th International Cosmic Ray Conference (ICRC2019). {\em arXiv} {\bf 2019}, arXiv:1908.05450. (accessed on 30-09-2021).
%Available online: \url{https://arxiv.org/abs/1908.05450v1} (\hl{accessed on}).

%Ref 63
\bibitem[Kimura(2018)]{Kimura:2018vvz} Kimura, S.S.; Murase, K.; Bartos, I.; Ioka, K.; Heng, I.S.; M\'{e}sz\'{a}ros, P. Transejecta high-energy neutrino emission from binary neutron star mergers. {\em Phys. Rev. D} {\bf 2018}, {\em 98}, 043020, {https://doi.org/10.1103/PhysRevD.98.043020}.

%Reference 64
\bibitem[Palacios(2021)]{Palacios:2021} Gonz\'{a}lez, J.P.; et al. [KM3NeT Collaboration]. KM3NeT/ARCA sensitivity to transient neutrino sources. In { Proceedings of the International Cosmic Ray Conference (ICRC2021), Berlin, Germany, 12--23 July 2021}; %Publisher: \hl{Italy,} {2021}; 
{Volume 395}, p. 1126.
%\url{https://pos.sissa.it/395/1126}.

% Reference 65
\bibitem[Oikonomou(2021)]{Oikonomou:2021} Oikonomou, F. High-energy neutrino emission from blazars. In {Proceedings of the International Cosmic Ray Conference (ICRC2021)}, Berlin, Germany, 12--23 July 2021; %Publisher: \hl{Italy}, {2021}, 
{Volume 395}, p. 030.
%\url{https://pos.sissa.it/395/030}.

% Reference 66
\bibitem[Fermi-LAT(2014)]{Fermi-LAT:2014ryh} Ackermann, M.;~Ajello, M.; Albert, A.; Atwood, W.B.; Baldini, L.; Ballet, J.; Barbiellini, G.; Bastieri, D.; Bechtol, K.; Bellazzini, R.; et~al. The spectrum of isotropic diffuse gamma-ray emission between 100 MeV and 820 GeV. {\em Astrophys. J.} {\bf 2015}, {\em 799}, 86, {https://doi.org/10.1088/0004-637X/799/1/86}.

%Ref 67
\bibitem[IceCube(2020)]{IceCube:2020wum} Abbasi, R.~et~al. [IceCube Collaboration]. The IceCube high-energy starting event sample: Description and flux characterization with 7.5 years of data. {\em Phys. Rev. D} {\bf 2021}, {\em 104}, 022002, {https://doi.org/10.1103/PhysRevD.104.022002}.

%Ref 68
\bibitem[Auger(2017)]{Auger:2017} Fenu, F.~et~al. [Pierre Auger Collaboration]. The cosmic ray energy spectrum measured using the Pierre Auger Observatory. In { Proceedings of the International Cosmic Ray Conference (ICRC2017), Busan, Korea, 12--20 July 2017}; %\Publisher: \hl{Italy}, { 2017}, 
{Volume 301}, p. 486.
%\url{https://pos.sissa.it/301/486}.

% Reference 69
\bibitem[Roth(2021)]{Roth:2021lvk} Roth, M.A.;~Krumholz, M.R.; Crocker, R.M.; Celli, S. The diffuse \ensuremath{\gamma}-ray background is dominated by star-forming galaxies. {\em Nature} {\bf 2021}, {\em 597}, 341--344 , {https://doi.org/10.1038/s41586-021-03802-x}.

% Reference 70
\bibitem[PierreAuger(2018)]{PierreAuger:2018qvk} Aab, A.~et~al. [The Pierre Auger Collaboration]. An Indication of anisotropy in arrival directions of ultra-high-energy cosmic rays through comparison to the flux pattern of extragalactic gamma-ray sources. {\em Astrophys. J. Lett.} {\bf 2018}, {\em 853},  L29, {https://doi.org/10.3847/2041-8213/aaa66d}.

%Ref 71
\bibitem[Lunardini(2019)]{Lunardini:2019zcf} Lunardini, C.; Vance, G.S.; Emig, K.L.; Windhorst, R.A. Are starburst galaxies a common source of high energy neutrinos and cosmic rays? {\em  J. Cosmol. Astropart. Phys.} {\bf 2019}, {\em 2019}, 073, {https://doi.org/10.1088/1475-7516/2019/10/073}.

%Ref 72
\bibitem[Tibet(2021)]{TibetASgamma:2021tpz} Amenomori, M.; et al. [Tibet $AS_\gamma$ Collaboration]. First Detection of sub-PeV Diffuse Gamma Rays from the Galactic Disk: Evidence for Ubiquitous Galactic Cosmic Rays beyond PeV Energies. {\em Phys. Rev. Lett.} {\bf 2021}, {\em 126}, 14, 141101, {https://doi.org/10.1103/PhysRevLett.126.141101}.

%Ref 73
\bibitem[Fang(2021)]{Fang:2021ylv} Fang, K.; Murase, K. Multi-messenger Implications of Sub-PeV Diffuse Galactic Gamma-Ray Emission. {\em Astrophys. J.} {\bf 2021}, {\em 919}, 93, {https://doi.org/10.3847/1538-4357/ac11f0}.

%Ref 74
\bibitem[HAWC(2019)]{HAWC:2019tcx} Abeysekara, A.U. et al. [HAWC Collaboration]. Multiple Galactic Sources with Emission Above 56 TeV Detected by HAWC. {\em Phys. Rev. Lett.} {\bf 2020}, {\em 124}, 021102, {https://doi.org/10.1103/PhysRevLett.124.021102}.

%Ref 75
\bibitem[Cao(2021)]{Cao:2021} Cao, Z.; Aharonian, F.A.; An, Q.; Bai, L.X.; Bai, Y.X.; Bao, Y.W.; Bastieri, D.; Bi, X.J.; Bi, Y.J.; Cai, H.; et al. Ultrahigh-energy photons up to 1.4 petaelectronvolts from 12 $\gamma$-ray Galactic sources. {\em Nature} {\bf 2021}, {\em 594}, 33--36, {https://doi.org/10.1038/s41586-021-03498-z}.

%Ref 76
\bibitem[Carpet(2021)]{Carpet-3Group:2021ygp} Dzhappuev, D.D. et al. [Carpet--3 Group]. Observation of Photons above 300 TeV Associated with a High-energy Neutrino from the Cygnus Region. {\em Astrophys. J. Lett.} {\bf 2021}, {\em 916}, 2, L22, {https://doi.org/10.3847/2041-8213/ac14b2}.


%%%%%%%%%%%%
% Reference 2
%\bibitem[Author2(year)]{ref-book1}
%Author~2, L. The title of the cited contribution. In {\em The Book Title}; Editor1, F., Editor2, A., Eds.; Publishing House: City, Country, 2007; pp. 32--58.

% Reference 3
%\bibitem[Author3(year)]{ref-book2}
%Author 1, A.; Author 2, B. \textit{Book Title}, 3rd ed.; Publisher: Publisher Location, Country, 2008; pp. 154--196.

% Reference 4
%\bibitem[Author4(year)]{ref-unpublish}
%Author 1, A.B.; Author 2, C. Title of Unpublished Work. \textit{Abbreviated Journal Name} stage of publication (under review; accepted; in~press).

% Reference 5
%\bibitem[Author5(year)]{ref-communication}
%Author 1, A.B. (University, City, State, Country); Author 2, C. (Institute, City, State, Country). Personal communication, 2012.

% Reference 6
%\bibitem[Author6(year)]{ref-proceeding}
%Author 1, A.B.; Author 2, C.D.; Author 3, E.F. Title of Presentation. In Title of the Collected Work (if available), Proceedings of the Name of the Conference, Location of Conference, Country, Date of Conference; Editor 1, Editor 2, Eds. (if available); Publisher: City, Country, Year (if available); Abstract Number (optional), Pagination (optional).

% Reference 7
%\bibitem[Author7(year)]{ref-thesis}
%Author 1, A.B. Title of Thesis. Level of Thesis, Degree-Granting University, Location of University, Date of Completion.

% Reference 8
%\bibitem[Author8(year)]{ref-url}
%Title of Site. Available online: URL (accessed on Day Month Year).

\end{thebibliography}

%=====================================
% References, variant B: internal bibliography
%=====================================

%%%%%%%%%%%%%%%%%%%%%%%%%%%%%%%%%%%%%%%%%%
%% for journal Sci
%\reviewreports{\\
%Reviewer 1 comments and authors’ response\\
%Reviewer 2 comments and authors’ response\\
%Reviewer 3 comments and authors’ response
%}
%%%%%%%%%%%%%%%%%%%%%%%%%%%%%%%%%%%%%%%%%%
\end{document}